\documentclass[12pt]{iopart}

\usepackage{iopams}
\usepackage{graphicx}
\input{epsf}
\begin{document}
\title[Polaron physics and crossover transition in magnetite]{Polaron physics and crossover transition in magnetite probed by pressure-dependent infrared spectroscopy}
\author{J. Ebad-Allah$^{1}$, L. Baldassarre$^{1}$, M. Sing$^{2}$, R. Claessen$^{2}$, V. A. M. Brabers$^{3}$, and C. A. Kuntscher$^{1}$}
\address{$^{1}$Experimentalphysik 2, Universit\"at Augsburg, D-86135
Augsburg, Germany}
\address{$^{2}$Physikalisches Institut, Universit\"at W\"urzburg, D-97074 W\"urzburg, Germany}
\address{$^{3}$Department of Physics, Eindhoven University of Technology, 5600 MB Eindhoven, The Netherlands}
\ead{christine.kuntscher@physik.uni-augsburg.de}

\begin{abstract}
The optical properties of magnetite at room temperature were studied
by infrared reflectivity measurements as a function of pressure up
to 8 GPa. The optical conductivity spectrum consists of a Drude term,
two sharp phonon modes, a far-infrared band at around 600~cm$^{-1}$, and a pronounced
mid-infrared absorption band. With increasing pressure both absorption bands
shift to lower frequencies and the phonon modes harden in a linear fashion.
Based on the shape of the MIR band, the temperature dependence of the dc transport data, and
the occurrence of the far-infrared band in the optical conductivity spectrum the polaronic coupling strength in
magnetite at room temperature should be classified as intermediate.
For the lower-energy phonon mode an abrupt increase of the linear pressure coefficient
occurs at around 6~GPa, which could be attributed to minor alterations of the charge distribution
among the different Fe sites.
\end{abstract}

\pacs{78.30.-j,62.50.-p,73.20.Mf}
\maketitle

\section{Introduction}

Magnetite (Fe$_{3}$O$_{4}$) belongs to the major class of highly
correlated electron systems and shows many interesting physical properties,
which triggered a large number of experimental and theoretical investigations over
the last 60 years. Fe$_{3}$O$_{4}$ has an inverse cubic spinel
structure at ambient conditions and is in a mixed-valence state
described as
[Fe$^{3+}$]$_{\mathrm{A}}$[Fe$^{2+}$+Fe$^{3+}$]$_{\mathrm{B}}$O$_{4}$,
where A and B denote the tetrahedral and octahedral sites,
respectively, in the spinel structure
AB$_{2}$O$_{4}$, with space group
Fd$\overline{3}$m \cite{Sasaki97}. Most of the proposed conduction mechanisms are
based on either band or polaron hopping motion of the extra electron
through the octahedral sites occupied by Fe$^{2+}$and Fe$^{3+}$
ions \cite{Ihle85,Schrupp05,Park98}. Magnetite undergoes a Verwey transition at
T$_\mathrm{v}$$\approx$120 K at ambient pressure towards an insulating state \cite{Verwey39}.
On cooling through the transition temperature, the dc conductivity
drops by two orders of magnitude concurrent with a first-order
structural phase transition from cubic to monoclinic
symmetry \cite{Iizumi82,Wright01,Wright02,Senn12}. The presence of the
mixed-valent Fe ions on the B sites motivated Verwey and
others \cite{Reviews80} to hypothesize that above T$_\mathrm{v}$
electronic exchange takes place whereas below T$_{\mathrm{v}}$
the transition is related to charge ordering (CO) in the
mixed-valent oxide. Recent resonant x-ray diffraction experiments have
specified the charge disproportionation  for T$<$T$_{\mathrm{v}}$ as
[Fe$^{3+}$]$_{\mathrm{A}}$[Fe$^{2.5+{\delta}}$+Fe$^{2.5-{\delta}}$]$_{\mathrm{B}}$O$_{4}$,
with $\delta_{12}$= 0.12$\pm$0.025 for one kind of Fe atom and
$\delta_{34}$= 0.1$\pm$0.06 for another kind \cite{Nazarenko06,Rozenberg06,Pasternak03,Rozenberg07}.

On the contrary, based on M\"ossbauer and powder x-ray diffraction
studies \cite{Rozenberg06,Kobayashi06,Pasternak03,Rozenberg07} it was claimed that
while cooling down at ambient pressure
a coordination crossover (CC) takes place at around T$_{\mathrm{v}}$,
where the spinel structure changes from inverse to normal:
\begin{equation} \label{coordination crossover}
[Fe^{3+}]_A[Fe^{2.5+}+Fe^{2.5+}]_BO_4  \rightarrow
[Fe^{2+}]_A[Fe^{3+}+Fe^{3+}]_BO_4.
\end{equation}
Within this scenario, $T_\mathrm{CC}$ increases with increasing pressure and reaches room temperature
at $\approx$6~GPa, while T$_{\mathrm{v}}$ decreases with increasing pressure \cite{Pasternak03,Rozenberg07}.
Concomitant with the CC transition a small change in the pressure dependence of the tetrahedral
and octahedral volumes was observed at around 6~GPa at room temperature, but no resolvable change
in the spinel-type crystal structure or unit cell volume \cite{Rozenberg07,Haavik00,Kuriki02}.
Furthermore, the hyperfine
interaction parameters as a function of pressure show an anomaly at $\approx$7~GPa, associated
with a pressure-induced discontinuous change of the Fe-O bond length \cite{Kobayashi06}.
The proposed valence transition from inverse to normal type is sluggish in character and hence spread
over the pressure range 7-15~GPa at room temperature \cite{Rozenberg07}.

The proposal of the occurrence of a CC transition in magnetite at room temperature has been recently discussed based on a
pressure-dependent thermoelectric power study combined with electrical resistance and sample's
contraction measurements \cite{Ovsyannikov08} and a high-pressure x-ray magnetic circular dichroism
study \cite{Baudelet10}. The results of both investigations exclude an inverse-to-normal spinel transition up to 20~GPa.
Furthermore, in Ref. \cite{Ovsyannikov08} a new pressure-induced crossover near 6~GPa was proposed,
with either a pressure-induced `perfection' of the electronic transport via mobile polarons, or a
pressure-induced driving of the dominant inverse spinel configuration to an `ideal' inverse spinel.


In the present work we carried out infrared reflectivity measurements to study
the electronic and vibrational properties of magnetite as a function
of pressure at room temperature. The goal of our study was two-fold:
first, to confirm the polaronic character of the electronic transport by
monitoring the typical spectral signatures of polarons in the infrared frequency range as a function
of pressure, and second, to test the proposed scenarios for the crossover
transition at $\sim$6~GPa based on the changes in the spectral features at around this
critical pressure.

\begin{figure}
\begin{center}
\includegraphics[width=0.5\columnwidth]{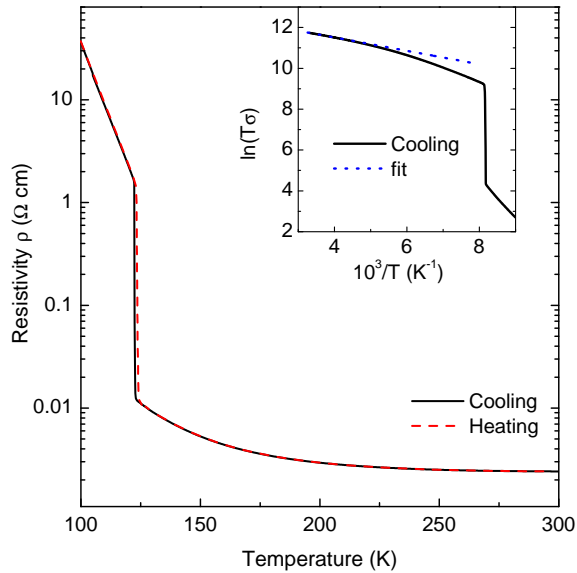}
\caption {Dc resistivity $\rho$ as a function of temperature. Inset: Arrhenius plot and fit (dotted line) of ln($T \sigma$) between 250 and
300 K according to Eq.\ (\ref{Eq:4}) for small polaron
hopping.}\label{fig:dc-data-3}
\end{center}
\end{figure}

\section{Experiment}

Single crystals of magnetite used in this work were prepared
from $\alpha$-Fe$_{2}$O$_{3}$ using a floating-zone method with
radiation heating \cite{Brabers71}. The quality of the crystals
was checked by electrical
transport measurements, showing a sharp increase of the resistivity
by a factor of about 100 at the critical temperature T$_\mathrm{v}\approx$
122 K, which is characteristic for the Verwey transition (see
Fig.~\ref{fig:dc-data-3}). Furthermore, the room temperature value of $\rho$ compares
well with earlier reports \cite{Miles57,Park98,Leonov07}.

In the pressure-dependent studies a clamp diamond anvil cell (Diacell cryoDAC-Mega)
equipped with type IIA diamonds, which
are suitable for infrared measurements, was used for the generation
of pressures up to 8 GPa. Finely ground CsI powder was chosen as
quasi-hydrostatic pressure medium. The single crystal was polished
to a thickness of $\approx$50 $\mu$m and a small piece (about 140
$\mu$m $\times$ 140 $\mu$m) was cut for each pressure measurement
and placed in the hole of the CuBe gasket. The pressure in the diamond anvil cell
was determined by the ruby luminescence method \cite{Mao86}.

Pressure-dependent reflectance measurements were conducted at room temperature (RT)
from 200 cm$^{-1}$ to 18000 cm$^{-1}$ using a Bruker IFS 66v/S
Fourier transform infrared (FT-IR) spectrometer. To focus the beam
on the small sample in the pressure cell, an infrared microscope
(Bruker IRscope II) coupled to the spectrometer and equipped with
a 15x magnification objective was used. The optical measurements
were partly carried out at the infrared beamline of the synchrotron
radiation source ANKA, where the same equipment is installed.
Reflectance spectra were measured at the interface between the
sample and the diamond anvil. Spectra taken at the gasket-diamond
interface served as the reference for
normalization of the sample spectra. Variations in the synchrotron source
intensity were taken into account by applying additional
normalization procedures. The reflectivity spectra are affected
at around 2000~cm$^{-1}$ by multiphonon absorptions in the diamond
anvils, which are not completely corrected by the normalization
procedure. This part of the spectrum was interpolated based on
the Drude-Lorentz fitting.
All reflectance spectra shown in this
paper refer to the absolute reflectance at the sample-diamond
interface, denoted as R$_{\mathrm{s-d}}$. The reflectance
spectrum of the free-standing sample (not shown) was found to be in
good agreement with earlier results \cite{Gasparov00,Park98}.

The real part of the optical conductivity, which directly shows the
induced excitations in a material, was obtained via Kramers-Kronig
(KK) transformation: To this end the measured reflectivity data were
extrapolated to low frequencies with a Drude-Lorentz fit, while data
from Ref. \cite{Park98} were merged at high frequencies. Moreover, the
sample-diamond interface was taken into account when performing the
KK analysis, as described in Ref.\cite{Pashkin06}.

\section{RESULTS}\label{section results}

\begin{figure}[t]
\begin{center}
\includegraphics[width=0.5\columnwidth]{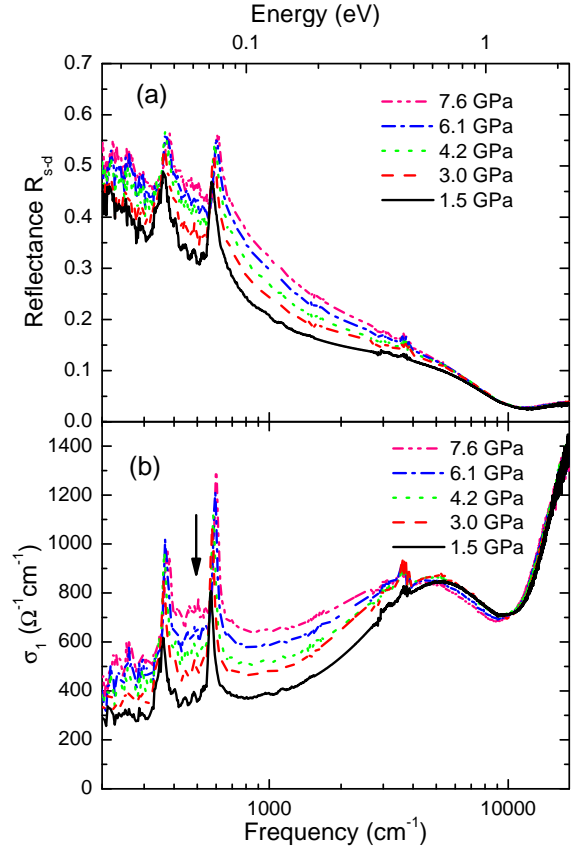}
\caption{(a) Pressure-dependent reflectance at RT and (b)
pressure-dependent real part of the optical conductivity at RT obtained by KK analysis.
The arrow in (b) marks the far-infrared band as discussed in the text.
The small features at around 3600~cm$^{-1}$ are artifacts due to multi-phonon
absorptions in the diamond anvil.}\label{fig:R-C-room}
\end{center}
\end{figure}

\begin{figure}[t]
\begin{center}
\includegraphics[width=0.5\columnwidth]{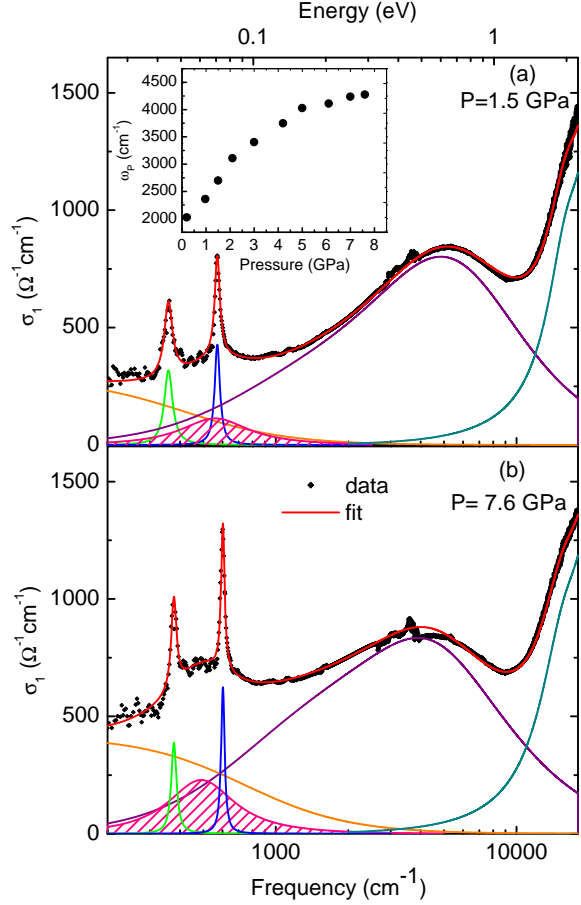}
\caption{Fit of the RT optical conductivity $\sigma_{1}$
at (a) 1.5 and (b) 7.6 GPa using the Drude-Lorentz model. The far-infrared band is highlighted by hatching. Inset: Plasma frequency $\omega_P$
of the Drude term as a function of pressure obtained from the Drude-Lorentz fitting.}
\label{fig:parameter}
\end{center}
\end{figure}

\begin{figure}[t]
\begin{center}
\includegraphics[width=0.5\columnwidth]{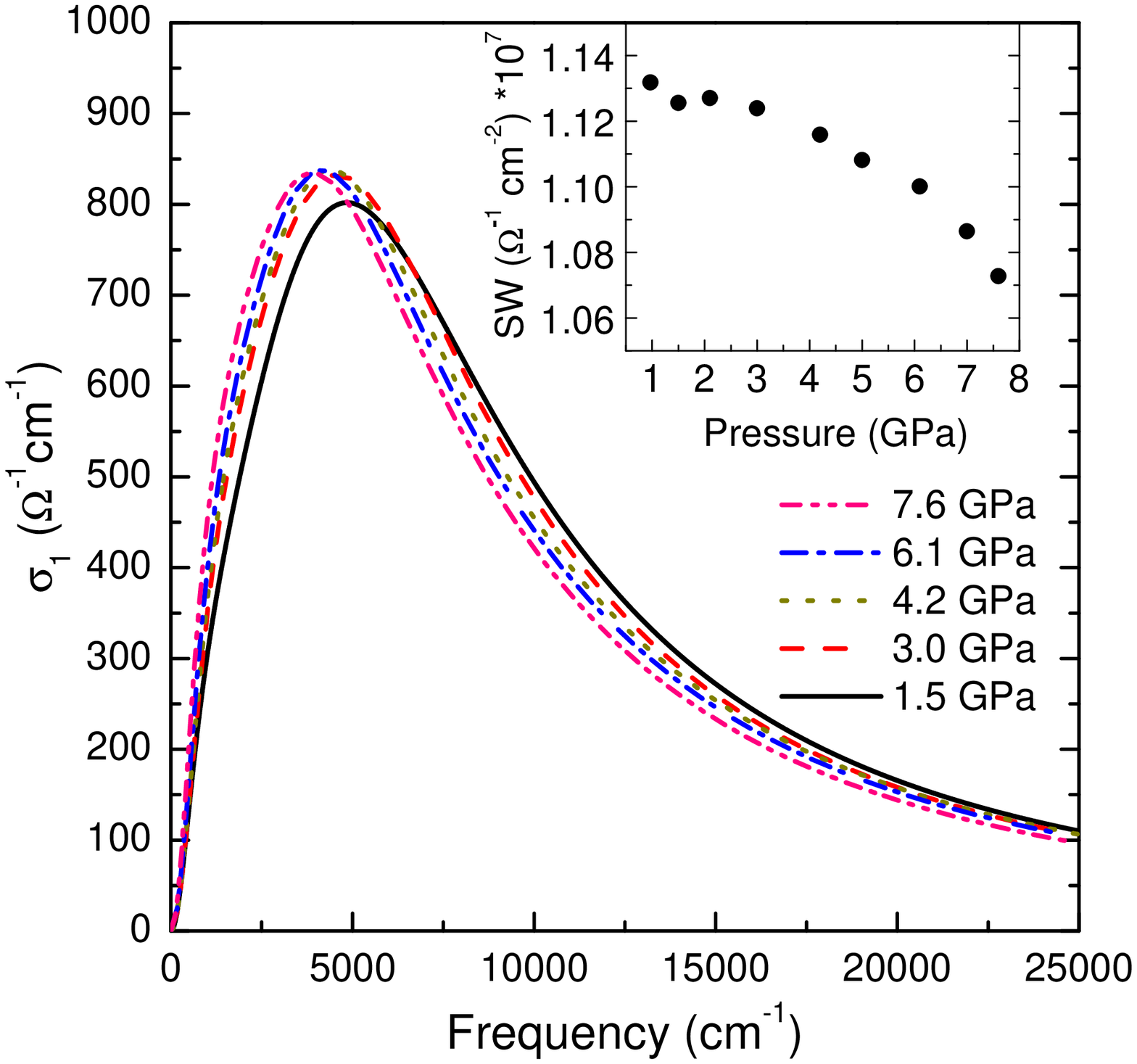}
\caption{MIR band as a function of pressure obtained by
Drude-Lorentz fitting of the optical conductivity spectra. Inset: Spectral weight
of the MIR band as a function of pressure.}\label{fig:mir-band}
\end{center}
\end{figure}

The infrared reflectance spectrum and the corresponding real part of the optical
conductivity of Fe$_{3}$O$_{4}$ as a function of pressure at RT are presented in
Fig.~\ref{fig:R-C-room}.
With increasing pressure the reflectance increases monotonically
for frequencies below 5000 cm$^{-1}$ ($\approx0.6$ eV), indicating
a growth of spectral weight within this frequency range. In contrast,
the reflectance is basically pressure independent above
12000 cm$^{-1}$($\approx1.5$ eV) \cite{Ebad-Allah09}.
We observe two T$_{1u}$ oxygen phonon modes near 355 cm$^{-1}$ and 565
cm$^{-1}$, consistent with earlier reports \cite{Gasparov00}.
The higher-frequency mode is attributed to the stretching of the A-O bond, whereas
the lower-frequency mode is related to the motion of the oxygen perpendicular
to this bond \cite{Waldron66,Brabers69}.
The real part of the optical conductivity of Fe$_{3}$O$_{4}$ is
presented in Fig.~\ref{fig:R-C-room} (b) for various pressures. The
finite conductivity in the far-infrared range suggests a Drude-type
contribution, indicating a metallic-like character of magnetite already at low pressure.
There is a pronounced mid-infrared (MIR) absorption band
located at around 5000 cm$^{-1}$($\approx0.6$ eV), which is a
typical signature of polaronic excitations \cite{Kuntscher03,Frank06,Thirunavukkuarasu06,Kuntscher06,Phuoc09} and
was attributed to the induced hopping of polaronic charge carriers
between the Fe$^{2+}$ and Fe$^{3+}$ ions on the octahedral
sites in magnetite \cite{Park98}. An additional far-infrared band is observed in the
optical conductivity spectrum at around 600~cm$^{-1}$, whose nature
is discussed in section~\ref{section:polaronic}.

To better quantify the changes of the optical conductivity with
increasing pressure, we fitted the conducti\-vi\-ty spectra with the
Drude-Lorentz model. The MIR band was modeled by two Lorentzian
functions and the far-infrared band at around 600~cm$^{-1}$ was described by
one Lorentzian oscillator.
As examples, we show in Fig.\ \ref{fig:parameter} the fitting
of the optical conductivity $\sigma_1$ at P = 1.5 and 7.6 GPa including the various
contributions, i.e., Drude term, two phonon modes, the far-infrared band, the MIR band, and higher-energy
interband transitions. This way we were able to extract the contributions
for each pressure applied.

With increasing pressure the MIR absorption band slightly narrows and redshifts,
while its spectral weight [$SW = \int \sigma_{1}(\omega') \mathrm{d}\omega'$] decreases with
increasing pressure (see inset of Fig.~\ref{fig:mir-band}). The
observed redshift confirms the polaronic nature of the band:
According to polaron theory \cite{Emin93} the frequency of the
polaron band is a measure of the polaron binding energy and thus of
the electron-phonon coupling. In general, the electron-phonon
coupling tends to decrease under pressure as a result of the
combined band broadening and stiffening of the crystal lattice.
Hence, one expects a decrease of the polaron binding energy under
pressure. The pressure-induced redshift of the polaronic absorption band
was also observed for other transition metal oxides \cite{Kuntscher05,Frank06}.
The evolution of the far-infrared band positioned at around 600~cm$^{-1}$ with pressure is depicted
in Fig.~\ref{fig:new}: It shifts monotonically to lower energy and its spectral weight
increases, with a saturation above $\approx$6~GPa (see inset of Fig.~\ref{fig:new}).

The stiffening of the lattice under pressure, as mentioned above, is
demonstrated by the hardening of the two phonon modes (see
Fig.~\ref{fig:phonons-room}): Both modes shift linearly with increasing pressure. The corresponding linear
pressure coefficient $B$ was obtained by fitting the peak positions
with the function $\omega(P)=A+B\times P$, where $P$ is the applied pressure.
For the higher-energy mode we obtain $B$=4.9~cm$^{-1}$/GPa for the whole studied pressure
range. For the lower-energy mode the linear pressure coefficient is $B$=1.9~cm$^{-1}$/GPa
up to $\approx$6~GPa; at this pressure an abrupt enhancement of the pressure-induced
hardening occurs, with $B$=4.9~cm$^{-1}$/GPa for the pressure range 6-8~GPa.
Interestingly, at the same pressure ($\approx$6~GPa) an anomaly is found in the
pressure dependence of the tetrahedral and octahedral
volume \cite{Rozenberg07}, the hyperfine interaction
parameters \cite{Kobayashi06}, and the sample's
contraction \cite{Ovsyannikov08}. In addition, drastic changes are
observed in the pressure dependence of the thermopower and
electrical resistance near 6~GPa \cite{Ovsyannikov08}.
This issue is discussed in section \ref{section:crossover}.

\section{DISCUSSION}

\subsection{Polaronic excitations} \label{section:polaronic}

In order to substantiate the qualitative picture of the
polaronic transport in magnetite, we
have analyzed our optical conductivity data using various
theoretical polaron models. Since
the nature of the polaron is reflected by the shape of the MIR
absorption band \cite{Emin93}, we fitted the MIR band in magnetite
using (i) a large polaron (LP) model and (ii) small
polaron (SP) models.

\begin{figure}[t]
\begin{center}
\includegraphics[angle=0,width=0.5\columnwidth]{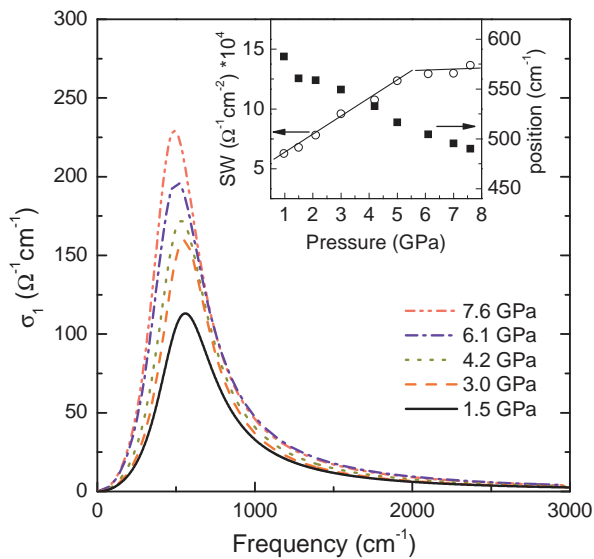}
\caption{Far-infrared band for selected pressures, obtained by
Drude-Lorentz fitting of the optical conductivity spectra.
Inset: Position and spectral weight of the far-infrared band as a function of
pressure. Lines are guides to the eye.} \label{fig:new}
\end{center}
\end{figure}

\begin{figure}[t]
\begin{center}
\includegraphics[angle=0,width=0.5\columnwidth]{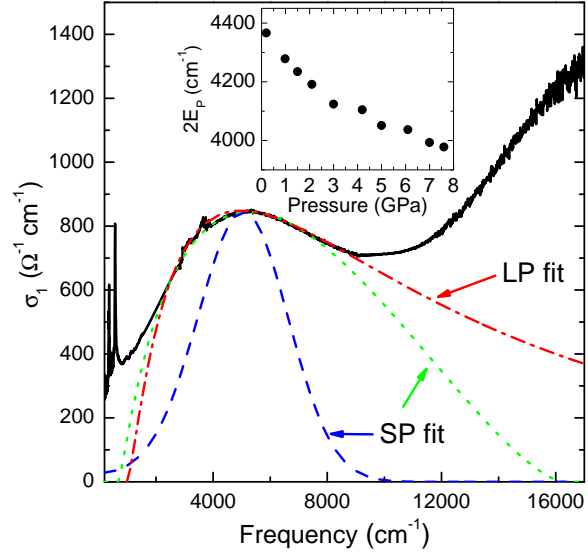}
\caption{Comparison between the fitting curves of the experimental optical conductivity data at 1.5~GPa using the LP model [Eq.
(\ref{Eq:1}), dashed-dotted line], the SP model in the limit $s\gg D$ [Eq. (\ref{Eq:2}),
dashed line], and the SP model in the limit $s\ll D$  [Eq. (\ref{Eq:3}), dotted line].
The inset displays the parameter 2E$_P$ as a function of
pressure, obtained from the fitting using Eq. (\ref{Eq:3}).} \label{fig:mir-polaron-LIS}
\end{center}
\end{figure}

(i) For large polarons the optical conductivity due to the
photoionization of the charge carriers as obtained by
Emin\cite{Emin93} is described by
\begin{equation}
\sigma_{1}(\omega)=n_{\mathrm{P}}\frac{64}{3}\frac{e^{2}}{\mathrm{m}}\frac{1}{\omega}\frac{(k(\omega)R)^{3}}{[1+(k(\omega)R)^{2}]^{4}},
\label{Eq:1}
\end{equation}
where $n_P$ is the polaron density and $R$ the polaron radius.
The photoionized carrier is treated as a free particle with mass $m$ and wavevector $k$.
The wave vector is defined as
$k=\sqrt{2m(\hbar\omega-3E_p)}/\hbar$, where $E_p$ is the polaron binding energy
or the ground state energy of the polaron. The threshold energy for the
photoionization of a large polaron is 3$E_p$ \cite{Emin93}.
By assuming
the effective mass for the large polaron to be equal to the free
electron mass $m_e$, the best fit of our data by using
Eq.~(\ref{Eq:1}) was achieved with the parameters
$n_P$ = 3.0$\times$ 10$^{21}$ cm$^{-3}$, $E_{p}$ = 327~cm$^{-1}$ (41~meV), and $R$ = 1.3~\AA. The LP
model describes the asymmetric lineshape of the MIR band
well, as illustrated in Fig.~\ref{fig:mir-polaron-LIS}.
Assuming that the conduction via polarons takes place among the octahedral B sites of the
spinel structure, we can estimate the effective number of charge carriers, $N_{eff}$, from $n_P$
according to $N_{eff}$=$n_P$$\times$$V/N$, where $N$=16 is the number of B-site atoms and
$V$=(8.394~\AA)$^3$ is the volume of the fcc unit cell \cite{Nazarenko06}. The resulting
$N_{eff}$=0.11 can be compared with the effective number of carriers per site obtained
from the spectral weight of the MIR band for LPs (see Fig.\ \ref{fig:mir-polaron-LIS}) according to
$N_{eff}=[2m_eV/(\pi e^2)]/N \int \sigma_1(\omega')d\omega'$,
where $m_e$ is the free electron mass, and $V$ the unit cell volume.
The so-obtained value $N_{eff}$=0.24 is about a factor of two higher than the value from
the fitting of the MIR band.
Furthermore, the value of the LP radius $R$ is smaller than the B--B distance,
$a_c$$\sqrt{2}$/4=2.97~\AA\ with the lattice constant $a_c$=8.394~\AA \cite{Hill79,Nazarenko06},
which is in disagreement with the LP scenario, where the LP radius should extend over multiple
lattice sites \cite{Emin93}.
This questions the applicability of the LP model for describing the optical data of magnetite.\footnote[1]{The discrepancies cannot be reconciled by considering a larger effective mass due to electronic correlations: For $m$ = 100 $\cdot$ $m_e$, as obtained in Ref.\cite{Gasparov00}, a good fit of the MIR band can only be obtained with the fitting parameters $n_P$ = 3.1 $\cdot$ 10$^{23}$ cm$^{-3}$, $E_{p}$ = 276~cm$^{-1}$, and $R$ = 0.1~\AA.}

\begin{figure}[t]
\begin{center}
\includegraphics[width=0.5\columnwidth]{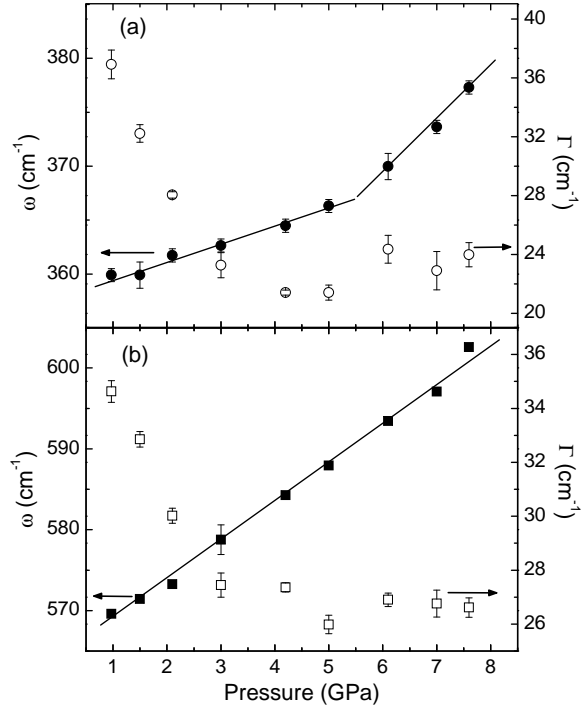}
\caption{Phonon frequencies and linewidths as a function of pressure
obtained by Drude-Lorentz fitting of the optical conductivity
spectra. The lines are linear fits to the data points, in order to
obtain the linear pressure coefficients.} \label{fig:phonons-room}
\end{center}
\end{figure}

(ii) The optical properties of small polarons have recently been reinvestigated
in the frame of the Holstein model using the dynamical mean-field theory \cite{Fratini06}.
In Ref.~\cite{Fratini06} both the antiadiabatic and the adiabatic
regimes for the polaron formation are discussed, which are distinguished by the
adiabaticity ratio $\gamma$=$\omega_0$/$D$, where $\omega_0$ is a typical phonon
frequency and $D$ is half the free-electron bandwidth.
We can estimate from our optical data the value of the adiabaticity ratio to $\gamma$$\approx$0.1
by using the theoretical value \cite{Piekarz07} $D$$\approx$0.7~eV (since we have $\omega_0$$\approx$75~meV).
Hence, we can refer to the formula in Ref.~\cite{Fratini06} obtained in the
adiabatic regime.

Furthermore, the phonon-induced broadening of the electronic levels, given by the
variance $s$ of the phonon field, compared to the electronic dispersion represented
by $D$, is crucial for the shape of the polaronic optical absorption \cite{Fratini06}.
The variance $s$ is given by \cite{Fratini06}
\begin{equation}
s^2(T)=E_P\omega_0 \coth \omega_0/(2T) .
\end{equation}
We estimate the value of $E_P$ from the position of the MIR band,
which is located at $\omega_{max}$$\approx$2$E_P$ for small polaron excitations,\cite{Emin93,Fratini06}
and thus obtain $E_P$$\approx$366~meV. With the typical phonon frequency of $\omega_0$$\approx$75~meV
for magnetite this gives $s$$\approx$165~meV at room temperature. Thus, the phonon-induced broadening is of the
same order of magnitude as the electronic band dispersion with $D$$\approx$700~meV \cite{Piekarz07}.
Therefore, we will consider both limits ($s$$\gg$$D$, $s$$\ll$$D$) in the following.

For $s$$\gg$$D$ the absorption by small polarons is described by a Gaussian-shaped peak centered at
$\omega_{max}$$\approx$2$E_P$ and is given at sufficiently high temperatures by the formula \cite{Reik63,Bi93a,Bi93b}
\begin{equation}
\sigma_{1}(\omega,\beta)=\sigma(0,\beta)\frac{\sinh(\frac{1}{2}\omega\beta)}{\frac{1}{2}\omega\beta}\exp[-\frac{\beta\omega^{2}}{16E_\mathrm{A}}],
\label{Eq:2}
\end{equation}
where $\beta = 1/(k_{B}T)$, and $E_\mathrm{A}$ is the thermal
activation energy, where $E_\mathrm{A}$=$E_{P}$/2. From the fitting we obtain
the dc conductivity $\sigma(0,\beta)$ = 34
$\Omega^{-1}$ cm$^{-1}$ and $E_\mathrm{A}$ = 1331~cm$^{-1}$ (165~meV). The value
of the thermal activation energy is consistent with earlier studies \cite{Gasparov00}.
However, the discrepancy between the fit and the measured data is rather large,
as seen in Fig.~\ref{fig:mir-polaron-LIS}.

For the other limit, $s$$\ll$$D$, the polaronic absorption band is calculated according to \cite{Fratini06}
\begin{equation}
\sigma_{1}(\omega)\propto\frac{1-e^{\omega/T}}{\omega}\Phi(\omega-2E_{\mathrm{P}})N(\omega-2E_{\mathrm{P}}),
\label{Eq:3}
\end{equation}
where the function $\Phi(\epsilon)$ is the corresponding current
vertex which can be derived by
$\Phi(\epsilon)=(D^{2}-\epsilon^{2})/3$, and the density of states
$N(\epsilon)=\frac{2}{\pi D^{2}}\sqrt{D^{2}-\epsilon^{2}}$ is
assumed to be semi-elliptical with the half-band width $D$. Within
this model the maximum of the absorption band is positioned at
$\omega_{max}=2E_P-D^2/2E_P$, i.e., it is shifted to lower energies relative to $2E_P$.
The expe\-ri\-mental data are well described by Eq.~(\ref{Eq:3}) (see Fig.~\ref{fig:mir-polaron-LIS}).
With increasing pressure the fitting parameter $2E_P$ is decreased,
as displayed in the inset of Fig.~\ref{fig:mir-polaron-LIS}, reflecting the pressure-induced shift
of the MIR band to lower frequencies.

According to Fig.~\ref{fig:mir-polaron-LIS} the SP model in the limit $s$$\ll$$D$
captures the shape of the MIR band best, since the fitting with the LP model gives a too small polaron radius.
A further test for the quasiparticles relevant for the transport is the
temperature dependence of the dc transport data above T$_\mathrm{v}$.
In case the conduction mechanism is due to the hopping of small polarons, one expects a thermally
activated behavior of the dc conductivity according to \cite{Mott71}
\begin{equation}
T\sigma(T)\propto\exp[-E_{H}/(k_{B}T)], \label{Eq:4}
\end{equation}
where $E_{H}$ is the hopping energy. Hereby, the disorder energy is omitted, which is
justified for crystalline bulk materials. The inset of Fig.\ \ref{fig:dc-data-3} displays the corresponding
Arrhenius plot. Obviously, the resulting curve does not follow a simple linear behavior.
By fitting the data in the temperature range 250 - 300~K according to Eq.\ (\ref{Eq:4})
one obtains a hopping energy $E_{H}$= 28.5~meV, which is a factor of $\approx$ 5 smaller
than the activation energy $E_{A}$=$E_{P}$/2 obtained from the optical data within the small polaron
model ($E_{P}$= 271~meV at the lowest pressure applied, see inset of Fig.~\ref{fig:mir-polaron-LIS}).
The discrepancy is, however, reduced
with decreasing temperature (see inset of Fig.\ \ref{fig:dc-data-3}),
indicating that the small polaron regime is gradually approached. Interestingly, recent hard x-ray
photoemission experiments \cite{Kimura10} also observed a gradual
variation from the large polaron to the small polaron regime with
decreasing temperature for the temperature range 250 - 330~K. We therefore conclude that
the polaronic coupling strength in magnetite is intermediate and the prevailing character
is temperature-dependent.

For this polaron crossover regime characteristic features are expected in the optical
conductivity spectrum, namely peaks at frequencies of the order of the phonon frequencies. These peaks correspond
to electronic transitions between different subbands in the polaron excitation spectrum, denoted as
polaron interband transitions \cite{Fratini01,Fratini06}.
Indeed, the optical conductivity spectrum contains a far-infrared band, which shifts to lower energy and
whose spectral weight is growing with increasing pressure (see Fig.\ \ref{fig:new}).
We speculate here about the polaronic nature of this far-infrared band.
The observations are si\-mi\-lar to those in other polaronic materials, like LaTiO$_{3.41}$
and $\beta$-Sr$_{1/6}$V$_{2}$O$_{5}$ \cite{Kuntscher03,Phuoc09},
which were also discussed within the intermediate electron-phonon coupling regime.

\subsection{Crossover at $\sim$6 GPa} \label{section:crossover}

Earlier pressure studies found an anomaly in the range 6-7~GPa in various physical
quantities. This anomaly was interpreted in terms of a crossover with a rearrangement of valence
electrons among the different types of Fe sites, including the scenario of an
inverse-to-normal spinel transition according to Eq.\ (\ref{coordination crossover}).
In the same pressure range ($\approx$6~GPa) we observe an enhancement of the pressure-induced
hardening of the lower-frequency phonon mode. Additional information about the pressure-induced changes
might be inferred from the linewidth of the phonon modes. In Ref.\ \cite{Gasparov00} an abrupt increase
of the linewidth of several phonon modes was observed while cooling through the Verwey transition. This increase
was interpreted in terms of small, non-resolvable splittings of the phonon modes due to crystal symmetry lowering.
In Fig.\ \ref{fig:phonons-room} we show, in addition to the frequencies, the linewidths of the phonon modes as
a function of pressure. For both modes the linewidth decreases as a function of pressure and is basically
pressure-independent above $\sim$4~GPa. Hence, we can exclude a pressure-induced symmetry lowering of the
crystal structure, since this would lift the degeneracy of the modes resulting in mode broadening \cite{Gasparov00}.

Overall, the pressure-induced changes in the optical conductivity spectrum are minor.
This observation is compatible with recent proposals of minor pressure-induced
alterations of the charge distribution in magnetite, without changing the principle transport mechanism via
polaron hopping motion -- e.g., a pressure-induced driving of the dominant inverse spinel configuration to an `ideal' inverse spinel \cite{Ovsyannikov08}. Interestingly, the parameters of several contributions to the optical response
saturate above $\sim$6~GPa (for example the plasma frequency of the Drude term or the spectral weight of the
far-infrared band), which is compatible with establishing an `ideal' inverse spinel configuration.

A similar pressure-induced rearrangement of charges between different transition-metal sites
was proposed recently for $\beta$-Na$_{0.33}$V$_2$O$_5$, as revealed by anomalies in the pressure-induced
shifts of infrared- and Raman-active phonon modes \cite{Kuntscher05,Frank07}. Similar to the findings
in magnetite, the charge rearrangement occurs without a change in the structural units and the
crystal structure symmetry \cite{Rabia09}.

\section{Summary}

In summary, the optical properties of magnetite have been studied at room temperature as a function of pressure,
supplemented by ambient-pressure dc transport measurements.
The optical conductivity spectrum consists of a Drude term,
two sharp phonon modes, a far-infrared band at $\approx$600~cm$^{-1}$, and a pronounced
MIR absorption band, which can be ascribed to excitations of polaronic quasiparticles.
With increasing pressure both absorption bands
shift to lower frequencies and the phonon modes harden in a linear fashion.
We applied several theoretical models for describing the MIR band and found
that the SP model in the limit $s$$\ll$$D$ captures the shape of the MIR band best.
However, according to the temperature dependence of the dc transport data
the small-polaron character of the charge carriers prevails only at low temperature,
and therefore the polaronic coupling strength in magnetite at room temperature should be rather
classified as intermediate.
This is corroborated by the occurrence of a far-infrared band in the optical conductivity spectrum.
For the lower-energy phonon mode an abrupt increase of the linear pressure coefficient
occurs at around 6~GPa, which can be attributed to minor alterations of the charge distribution
among the different Fe sites.

\section*{Acknowledgements}
We acknowledge the ANKA Angstr\"omquelle Karlsruhe for the provision
of beamtime and thank B. Gasharova, Y.-L. Mathis,
D. Moss, and M. S\"upfle for assistance using the beamline ANKA-IR.
Fruitful discussions with S. Fratini are greatfully acknowledged.
This work was financially supported by the DFG through SFB 484.

\end{document}